\documentclass[prd,aps,amsmath,amssymb,epsfig]{revtex4}
\usepackage{graphicx}
\usepackage{dcolumn}
\usepackage{bm}

\newcommand{\mn}{{\mu\nu}}
\newcommand{\half}{{1\over 2}}
\newcommand{\bea}{\begin{eqnarray}}
\newcommand{\eea}{\end{eqnarray}}
\begin{document}
\title{General Structure of Relativistic Vector Condensation}
\author{Francesco {\sc Sannino}}
\email{francesco.sannino@nbi.dk} \affiliation{{\rm NORDITA},
Blegdamsvej 17, DK-2100 Copenhagen \O, Denmark}

\begin{abstract}

We study relativistic massive vector condensation due to a non
zero chemical potential associated to some of the global conserved
charges of the theory. We show that the phase structure is very
rich. More specifically there are three distinct phases depending
on the value of one of the zero chemical potential vector self
interaction terms. We also develop a formalism which enables us to
investigate the vacuum structure and dispersion relations in the
spontaneously broken phase of the theory. We show that in a
certain limit of the couplings and for large chemical potential
the theory is not stable. This limit, interestingly, corresponds
to a gauge type limit often employed to economically describe the
ordinary vector mesons self interactions in QCD. We finally
indicate for which physical systems our analysis is relevant.
\end{abstract}

\maketitle

\section{Introduction}
\label{introduction}

Relativistic vector condensation has been proposed and studied in
different realms of theoretical physics. However the condensation
mechanism and the nature of the relativistic vector mesons
themselves is quite different. Linde \cite{Linde:1979pr}, for
example, proposed the condensation of the intermediate vector
boson $W$ in the presence of a superdense fermionic matter while
Ambj\o rn and Olesen \cite{Ambjorn:1989gb} investigated their
condensation in presence of a high external magnetic field. Manton
\cite{Manton:1979kb} and later on Hosotani \cite{Hosotani:1988bm}
considered the extension of gauge theories in extra dimensions and
suggested that when the extra dimensions are non simply connected
the gauge fields might condense. Li in \cite{Li:2002iw} has also
explored a simple effective Lagrangian and the effects of vector
condensation when the vectors live in extra space
dimensions\footnote{The effective Lagrangian and the condensation
phenomenon in extra (non compact) space dimensions has also been
studied/suggested in \cite{Sannino:2001fd}.}.

Brown and Rho \cite{Brown:kk} also suggested that the non gauge
vectors fields such as the (quark) composite field $\rho$ in QCD
may become light and possibly condense in a high quark matter
density and/or in hot QCD. Harada and Yamawaki's
\cite{Harada:2000kb} dynamical computations within the framework
of the hidden local gauge \cite{BKY} symmetry support this
picture.

Rotational symmetry can also break in a color superconductor if
two quarks of the same flavor gap. In this case the quarks must
pair in a spin one state and a careful analysis has been performed
in \cite{Pisarski:1999tv}. Whether this gap occurs or not in
practice is a dynamical issue recently investigated in
\cite{Buballa:2002wy}.

We consider another type of condensation. If vectors themselves
carry some global charges we can introduce a non zero chemical
potential associated to some of these charges. If the chemical
potential is sufficiently high one can show that the gaps (i.e.
the energy at zero momentum) of these vectors become light
\cite{Lenaghan:2001sd,{Sannino:2001fd}} and eventually zero
signaling an instability. If one applies our results to 2 color
Quantum Chromo Dynamics (QCD) at non zero baryon chemical
potential one predicts that the vectors made out of two quarks (in
the 2 color theory the baryonic degrees of freedom are bosons)
condense. Recently lattice studies for 2 color at high baryonic
potential \cite{Alles:2002st} seem to support our predictions.
This is the relativistic vectorial Bose-Einstein condensation
phenomenon. A decrease in the gap of vectors is also suggested at
high baryon chemical potential for two colors in
\cite{Muroya:2002ry}.

Interestingly non-relativistic vectorial Bose-Einstein
condensation recently has attracted much attention in condensed
matter physics since it has been observed experimentally in alkali
atom gases \cite{VBE}. We also note that in this framework a rich
phenomenology related to the classical solutions of the theory,
such as vortices \cite{Volovik:2000mt}, is expected to occur.

Here we extend the analysis presented in
\cite{Lenaghan:2001sd,{Sannino:2001fd}} by showing the existence
of new phases while providing a detailed investigation of the
relativistic massive vector condensation phenomenon driven by a
non zero chemical potential. We develop an efficient formalism
which allows us to analytically study the dispersion relations of
the vectors in the broken phase of the theory and set the stage
for possible higher order computations. Our Lagrangian approach
has to be understood as a Landau theory describing the vector
condensation phenomenon.

Having in mind some specific physical applications we assume our
massive relativistic vectors to be in the adjoint representation
of the non abelian $SU(2)$ global symmetry group. We then turn on
a chemical potential in a specific $SU(2)$ direction which breaks
$SU(2)$ explicitly to a $U(1)$ symmetry. We demonstrate that we
can have three independent phases according to the value assumed
by one of the two distinct (non derivative) vector self
interactions. The polar phase, the apolar phase and the enhanced
symmetry one. The polar phase has been introduced in
\cite{Sannino:2001fd} along with the enhanced symmetry case. The
polar and apolar phases are the relativistic generalization of the
 phases encountered in the condense matter \cite{Volovik:2000mt}
framework. The appearance of a given phase is related to the
specific value of one of the vector self interaction coefficients
of the Lagrangian at zero chemical potential. In this paper we
study consistently the gapless excitations and the dispersion
relations in all of the phases while providing a formalism helpful
when trying to go beyond the tree level approximation.

In the polar phase the vector condensate breaks the rotational
symmetry down to a simple rotation around the axis of
condensation. The internal $U(1)$ symmetry breaks completely as
well. This phase is characterized by three gapless excitations
with linear in momentum dispersion relations. By studying in
detail the dispersion relations we show that two of the gapless
states have isotropic dispersion relations while the third
excitation has different velocities in the direction parallel and
orthogonal to the vacuum expectation value. The detailed analysis
of the dispersion relations is exact when we supplement the global
symmetry $SU(2)$ by an extra one which allows only an even number
of vectors in any vertex of the theory, however the number and
type of goldstone bosons is independent of the extra discrete
symmetry.

If we are in the apolar phase the condensation of the vector is
such that the unbroken generator is a linear combination of one of
the rotational generators and the abelian $U(1)$ generator. Only
two gapless excitations emerge this time. One with linear and the
other with quadratic dispersion relations. However the vacuum
still breaks three generators. This is in agreement with the
Nielsen and Chadha \cite{Nielsen} counting scheme according to
which in absence of Lorentz invariance (and under a number of
assumptions) each gapless excitation with quadratic dispersion
relations has to be counted as two goldstone bosons with linear
dispersion relations. More specifically if $n_I$ denotes the
number of gapless excitations of type $I$ with linear dispersion
relations (i.e. $E\propto p$) and $n_{II}$ the ones with quadratic
dispersion relations (i.e. $E\propto p^2$) the Goldstone theorem
generalizes to $n_I+2\,n_{II} \ge \#~{\rm broken~generators}$.
Clearly in the absence of type $II$ excitations we recover the
usual counting. We analyze the spectrum in this case and find that
the type II goldstone boson has isotropic dispersion relations
while the type I has not. In all of the phases the non goldstone
boson excitations are investigated in detail as well.

If one of the self-vector couplings is set to zero we observe
three gapless excitations: 2 type II goldstone bosons and one type
I. In this case the potential has an enhanced global symmetry,
i.e. it has an $SO(6)$ which the vacuum breaks to an $SO(5)$. In
absence of Lorentz breaking, we would have 5 ordinary goldstone
bosons. However the presence of the chemical potential in the
derivative terms of the Lagrangian prevents the emergence of 2
extra goldstone bosons \cite{Schafer:2001bq} while turning 2
goldstones into type II \cite{Sannino:2001fd}.

We also discover that when the vector self interaction couplings
are tuned to be the ones predicted using the Yang-Mills relation
(while keeping always a non zero vector mass) the theory at high
chemical potential does not predict a stable solution. The
existence of an inhomogeneous phase is not excluded. However a
more natural solution of the instability is that the chemical
potential, when increasing the associated charge density, is at
the most as large as the mass of the vectors.

If we now apply our results to describe non elementary
relativistic massive vectors such as the $\rho$ field of QCD at
high chemical potential (isospin for example) it will help
elucidating and constructing the effective Lagrangians for vector
fields at zero chemical potential. Interestingly, indeed, the
particular choice of the vector coupling respecting Yang-Mills
relations is the one often assumed in literature when introducing
the quark composite field $\rho$ at the effective Lagrangian level
\cite{KRS,{KS}}. {}For example the gauge choice of the couplings
emerges very naturally when the vector mesons are introduced as
gauge fields of an hidden local gauge symmetry \cite{BKY}. By
studying the vector condensation phenomenon (for example the
ordinary $\rho$ at high isospin chemical potential) on the lattice
it is possible to shed light on the correct way of constructing a
theory for composite vector fields or massive vector fields in
general. These effective theories for composite fields are
relevant also for the physics beyond the standard model of
particle interactions such as the ones relative to a strongly
interacting electroweak sector (see \cite{ARS,{DRS}}).

In section \ref{vacuum} we present the Lagrangian and study the
vacuum. We show that a number of phases can be present. In sec.
\ref{dispersion} we solve for the dispersion relations. We then
conclude while reviewing the physical applications in sec.
\ref{conclusion}. In the appendices we summarize first all of the
terms of the Lagrangian in the cartesian and cylindrical
coordinates. We finally show the computational details of the
dispersion relations for the apolar case.

\section{Vacuum Structure and Different Phases}
\label{vacuum} There are different ways to describe vector fields
at the effective Lagrangian level. {}For example one can use the
hidden local gauge symmetry of Ref.\ \cite{BKY}, or the
antisymmetric tensor field of Ref.\ \cite{Ecker:1993de}. Or one
can introduce the massive vector fields following the method
outlined in \cite{KRS,KS}. {}For most of the known physical
applications these approaches provide identical results (at the
tree level). It is worth noticing that in some of these methods
certain coefficients of the vector Lagrangian are related by
enforcing the
Yang-Mills relations. 

We choose to consider the following general effective Lagrangian
for a relativistic massive vector field in the adjoint of $SU(2)$
in $3+1$ dimensions and up to four vector fields, two derivatives
and containing only intrinsic positive parity terms\footnote{{}For
simplicity and in view of the possible physical applications we
take the vectors to belong to the adjoint representation of the
$SU(2)$ group.}:
\begin{eqnarray}
{\cal
L}&=&-\frac{1}{4}F^a_{\mu\nu}F^{a{\mu\nu}}+\frac{m^2}{2}A_{\mu}^a
A^{a\mu} +\delta\, \epsilon^{abc}\partial_{\mu} A_{a\nu} A^{\mu}_b
A^{\nu}_c  - \frac{\lambda}{4}\left(A^a_{\mu}A^{a{\mu}}\right)^2 +
 \frac{\lambda^{\prime}}{4} \left(A^a_{\mu}A^{a{\nu}}\right)^2 \ ,
 \end{eqnarray}
with $F_{\mu
\nu}^a=\partial_{\mu}A^a_{\nu}-\partial_{\nu}A^a_{\mu}$, $a=1,2,3$
and metric convention $\eta^{\mu \nu}={\rm diag}(+,-,-,-)$. Here,
$\delta$ is a real dimensionless coefficient, $m^2$ is the tree
level mass term and $\lambda$ and $\lambda^{\prime}$ are positive
dimensionless coefficients with $\lambda \ge \lambda^{\prime}$
when $\lambda^{\prime}\ge 0$ or $\lambda\ge 0$ when
$\lambda^{\prime}\ge 0$ to insure positivity of the potential. The
Lagrangian describes a self interacting $SU(2)$ Yang-Mills theory
in the limit $m^2=0$, $\lambda=\lambda^{\prime}>0$ and $
\delta=-\sqrt{\lambda}$.

It is relevant to notice that in the limit $\delta=0$ the theory
gains a new symmetry according to which we have always a total
number of even vectors in any process. This symmetry guarantees
that if the $\delta$ term is absent from the start it will not be
generated dynamically. In this paper we will mainly investigate
the theory in this case since it will simplify our computations.
However we will comment on the effects of such a term in a final
paragraph.

The effect of a nonzero chemical potential associated to a given
conserved charge - (say $\displaystyle{T^3=\frac{\tau^3}{2}}$) -
can be readily included \cite{Lenaghan:2001sd} by modifying the
derivatives acting on the vector fields:
\begin{equation}
\partial_{\nu} A_{\rho} \rightarrow \partial_{\nu}A_{\rho} - i
\left[B_{\nu}\ ,A_{\rho}\right]\ ,
\end{equation}
with $B_{\nu}=\mu \,\delta_{\nu 0} T^3\equiv V_{\nu} T^3$ where
$V=(\mu\ ,\vec{0})$. In appendix it is summarized the effective
Lagrangian, in this basis, after the introduction of the chemical
potential. The introduction of the chemical potential breaks
explicitly the Lorentz transformation leaving invariant the
rotational symmetry. Also the $SU(2)$ internal symmetry breaks to
a $U(1)$ symmetry. If the $\delta$ term is absent we have an extra
unbroken $Z_2$ symmetry which acts according to
$A^3_{\mu}\rightarrow -A^3_{\mu}$. These symmetries suggest
introducing the following cylindric coordinates:
\begin{eqnarray}
\phi_\mu &=& {1 \over \sqrt{2}} ( A^1_\mu + i A^2_\mu )\ , \qquad
\phi^{\ast}_\mu = {1 \over \sqrt{2}} ( A^1_\mu - i A^2_\mu )\ ,
\qquad \psi_\mu = A^3_\mu \ ,
\end{eqnarray}
on which the covariant derivative acts as follows:
\begin{eqnarray}
D_{\mu}\phi_{\nu} = \left(\partial + i\, V \right)_{\mu}
\phi_{\nu} \ , \quad  D_{\mu}\psi_{\nu} = \partial_{\mu}
\psi_{\nu} \ , \quad V_\nu = \left(\mu,{\bf 0}\right) \ .
\end{eqnarray}
The quadratic, cubic and quartic terms - in the vector fields - in
the cylindrical coordinates are summarized in the appendices.
\subsection{The Non Derivative Terms}
We first study the non derivative terms which we collect in the
following potential type term:
\begin{eqnarray}
V&=&- \frac{m^2}{2} \left[ 2{\phi^{\ast 0}} \phi^0 + \psi \cdot
\psi \right] + \left(m^2 - \mu^2 \right)\,\left[ {\phi^{\ast 0}}
\phi^0-\phi^{\ast}\cdot \phi \right]+{2\lambda - \lambda^{\prime}
\over 2} (\phi^{\ast} \cdot \phi)^2 - {\lambda^{\prime}\over 2}
|(\phi \cdot \phi)|^2 +
{\lambda- \lambda^{\prime} \over 4} (\psi \cdot \psi)^2 \nonumber \\
&& +{\lambda} (\phi^{\ast} \cdot \phi)(\psi \cdot \psi) -
\lambda^{\prime} |(\psi \cdot \phi)|^2 + 2\mu\delta\, \psi^0
\left(\phi^{\ast}\cdot
 \phi\right)+ \mu \delta\, \left[{\phi^{\ast}}^0 \left(
 \phi\cdot \psi\right) + {\rm  c.c.} \right] \ .
\end{eqnarray}
We can read off the symmetries of the theory from the potential.
When $\delta=0$, for example, we gain the discrete symmetry
$\psi\rightarrow -\psi$.  To explore the vacuum structure of the
theory we consider the following variational ansatz:
\begin{eqnarray}
\psi^{\mu}= 0\ , \qquad \phi^{\mu} = \sigma \,
\left(%
\begin{array}{c}
  0 \\
  1 \\
    e^{i\alpha}\\
  0 \\
\end{array}%
\right) \ .
\end{eqnarray}
Substituting the ansatz in the potential expression we have:
\begin{eqnarray}
V=2\, \sigma^4\left[(2\lambda - \lambda^{\prime})-
\lambda^{\prime}\cos^2\alpha \right] + 2\left(m^2 - \mu^2\right)
\sigma^2 \ .
\end{eqnarray}
The potential is positive for any value of $\alpha$ when $\lambda
> \lambda^{\prime}$ if $\lambda^{\prime}\ge 0$ or $\lambda >0$ if
$\lambda^{\prime}<0$. Due to our ansatz the ground state is
independent of $\delta$. The unbroken phase occurs when $\mu \leq
m$ and the minimum is at $\sigma=0$. A possible broken phase is
achieved when $\mu> m$ since in this case the quadratic term in
$\sigma$ is negative. According to the value of $\lambda^{\prime}$
we distinguish three distinct phases:
\subsection{The polar phase: $\lambda^{\prime}>0$}
In this phase the minimum is for
\begin{eqnarray}
 \langle \phi^{\mu} \rangle = \sigma \,
\left(%
\begin{array}{c}
  0 \\
  1 \\
  1\\
  0 \\
\end{array}%
\right) \ , \quad {\rm with} \quad \sigma^2=\frac{1}{4}\frac{\mu^2
- m^2}{\lambda - \lambda^{\prime} }\ ,
\end{eqnarray}
This phase has been partially analyzed in \cite{Sannino:2001fd}.
We have the following pattern of symmetry breaking $SO(3)\times
U(1) \rightarrow SO(2)$. If we define with ${R_i}$ the three
generators of $SO(3)$ (acting only on the spatial indices) as
follows:
\begin{eqnarray}
R_1=\left(
\begin{array}{ccc}
  0 & 1 & 0 \\
  -1 & 0& 0 \\
  0& 0 & 0 \\
\end{array}
\right) \ ,
\quad R_2=\left(%
\begin{array}{ccc}
  0 & 0 & 1 \\
  0 & 0& 0 \\
  -1& 0 & 0 \\\end{array} \right) \ ,
  \quad R_3=\left(%
\begin{array}{ccc}
  0 & 0 & 0 \\
  0 & 0& 1 \\
  0& -1 & 0 \\
\end{array}%
\right)\ ,
\end{eqnarray}
while the $U(1)$ generator acts as a phase on $\phi$ then the
unbroken generator determined imposing invariance of the vacuum
\begin{eqnarray}
T_{\rm unbroken}^{\rm polar} \langle \phi \rangle = 0\ ,
\end{eqnarray}
is the linear combination $T_{\rm unbroken}^{\rm polar}=R_2-R_3$.
We have three broken generators. We can show (see section on the
dispersion relations for details) that in this case we have 3
gapless excitations with linear dispersion relations. All of the
physical states (with and without a gap) are either vectors
(2-component) or scalars with respect to the unbroken $SO(2)$
group. The dispersion relations for the 3 gapless states
\cite{Sannino:2001fd} computed explicitly in the following
sections are reported here for the reader's convenience.
\begin{eqnarray}
E^2_{\Phi_V^{-}} &\propto& {\lambda^{\prime}} (\mu^2 - m^2) \,
{\bf p}^2  \ ,  \qquad \qquad{\rm~~~~~~~V=2~Physical~States} \label{first}\\
E_{{\Phi_S}^-}^2 &\propto& (\mu^2-m^2)\,\left(p^2_{\perp} +
{v}_{{\Phi_S}^-_{\parallel}}^2 p^2_{\parallel}\right) \ ,
\quad {\rm S=1~Physical~State} \\
{v}_{{\Phi_S}^-_{\parallel}}^2&\propto& \left(\lambda -
\lambda^{\prime} \right)\frac{\mu^2\,\lambda +
\lambda^{\prime}\left(\mu^2 - 2m^2\right)}{(\mu^2\lambda -m^2
\lambda^{\prime})^2} \ , \label{example}
\end{eqnarray}
where $p_{\parallel(\perp)}$ refers to the momentum  parallel
(orthogonal) to the vacuum. Here we present the leading terms in a
momentum expansion of the gapless dispersion relations. The vector
components orthogonal to the vacuum direction (2 states indicated
with $V$) propagate isotropically while the component in the
direction of the vacuum (1 state indicated with $S$) does not. As
a consistency check one sees that at $\mu=m$ the dispersion
relations are all isotropic. At $\mu=m$ the dispersion relations
are no longer linear in the momentum. This is related to the fact
that the specific part of the potential term has a partial
conformal symmetry discussed first in \cite{Sannino:2001fd}. Some
states in the theory are curvatureless but the chemical potential
term present in the derivative term prevents these states to be
gapless. There is a transfer of the conformal symmetry information
from the potential term to the vanishing of the velocity of the
gapless excitations related to the would be gapless states. This
conversion is due to the linear time-derivative term induced by
the presence of the chemical potential term
\cite{Schafer:2001bq,{Sannino:2001fd}}.


\subsection{Enhanced symmetry and type II Goldstone bosons: $\lambda^{\prime}=0$}
Here the potential has an enhanced $SO(6)$ in contrast to the
$SU(2)\times U(1)$ for $\lambda^{\prime}\neq 0$ global symmetry
which breaks to an $SO(5)$ with 5 broken generators. Expanding the
potential around the vacuum we find 5 null curvatures
\cite{Sannino:2001fd}. However we have only three gapless states
obtained diagonalizing the quadratic kinetic term and the
potential term (see the dispersion relation section). Two states
(a vector of $SO(2)$) become type II goldstone bosons (see
eq.~(\ref{first})) while the scalar state remains type I. This
latter state is the goldstone boson related to the spontaneously
broken $U(1)$ symmetry. According to the Nielsen-Chada theorem the
type II states are counted twice with respect to the number of
broken generators while the linear just once recovering the number
of generators broken by the vacuum. More specifically one can
prove \cite{Sannino:2001fd} that the velocity of the states
labelled by $V$ in eq.~(\ref{first}) is proportional to the
curvatures (evaluate on the minimum) of the would be goldstone
bosons which is zero in the $\lambda^{\prime}=0$ limit. Again we
have an efficient mechanism for communicating the information of
the extra broken symmetries from the curvatures to the velocities
of the already gapless excitations.
\subsection{The apolar phase: $\lambda' < 0$}
We now extend the vacuum analysis in \cite{Sannino:2001fd} to the
apolar phase in which the potential is minimized for:
\begin{eqnarray}
\langle \phi^{\mu} \rangle= \sigma \,
\left(%
\begin{array}{c}
  0 \\
  1 \\
  \imath\\
  0 \\
\end{array}%
\right) \ , \quad {\rm with} \quad\sigma^2 =\frac{1}{2}\frac{\mu^2
- m^2}{2\lambda - \lambda^{\prime}} \ .
\end{eqnarray}
In this phase we have again 3 broken generators. However the
unbroken generator is the following combination of the $U(1)$ and
$SO(3)$ generator:
\begin{eqnarray}
T_{\rm unbroken}^{\rm apolar}=R_1 - i\,{\bf 1} \ .
\end{eqnarray}
More explicitly the action on the vev is $(R_1-\imath\,{\bf
1})\langle \phi \rangle=0$. In this phase as we shall demonstrate
in the next sections we have only two gapless states. One of the
two states is a type I goldstone boson while the other is type II.
The two goldstone bosons are one in the $z$ and the other in the
$x-y$ plane. Interestingly in this phase, due to the intrinsic
complex nature of the vev, we have spontaneous $CP$ breaking.

\begin{figure}[h]
\includegraphics[width=7truecm, height=4truecm]{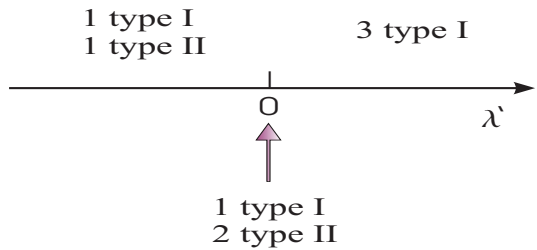}
\vskip -1cm \caption{We show the number and type of goldstone
bosons in the three distinct phases associated to the value
assumed by the coupling $\lambda^{\prime}$. In the polar phase,
positive $\lambda^{\prime}$, we have 3 type I goldstone bosons. In
the apolar phase, negative $\lambda^{\prime}$, we have one type I
and one type II goldstone boson while in the enhanced symmetry
case $\lambda^{\prime}=0$ we have one type I and two type II
excitations.} \label{Figura1}
\end{figure}

We summarize in Fig.~{\ref{Figura1}} the phase structure in terms
of the number of goldstone bosons and their type according to the
values assumed by $\lambda^{\prime}$.
\subsection{The case $\lambda = \lambda^{\prime}$: the gauge theory
limit} 
\begin{figure}[hbt]
\includegraphics[width=8truecm, height=6cm]{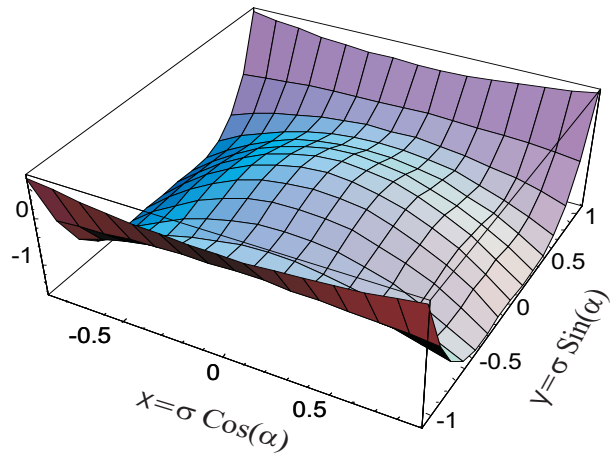}
\caption{Potential plotted for $\mu=2m$ and
$\lambda=\lambda^{\prime}=1$.} \label{gauge}
\end{figure}

Here the potential is:
\begin{eqnarray}
V=2\, \sigma^4\lambda \sin^2\alpha  + 2\left(m^2 - \mu^2\right)
\sigma^2 \ .
\end{eqnarray}
This potential has two extrema when $\mu>m$, one for $\alpha=0$
and $\sigma=0$ which is an unstable point and the other for
$\alpha=\pm\pi/2$ and $\sigma^2=\frac{(\mu^2-m^2)}{2\lambda}$
corresponding to a saddle point (see the potential in
Fig.~\ref{gauge}).
 At first the fact that we have no stable
solutions seems unreasonable since we know that in literature we
often encounter condensation of intermediate vector mesons such as
the $W$ boson. However (except for extending the theory in higher
space dimensions) in these cases one often introduces an external
source. {}For example one adds to the theory a strong magnetic
field (say in the direction $z$) which couples to the
electromagnetically charged intermediate vector bosons $W^{+}$ and
$W^{-}$). In this case the potential is (see
\cite{Ambjorn:1989gb}):
\begin{eqnarray}
V=2\, \sigma^4\lambda \sin^2\alpha  + 2\left(m^2 -
e\,H\,\sin\alpha\right) \sigma^2 \ ,
\end{eqnarray}
where $e$ is the electromagnetic charge and $H$ is the external
electromagnetic source field. This potential has a true minimum
for $\alpha=\pi/2$ and $\sigma=\frac{eH-m^2}{2\lambda}$ whenever
the external magnetic field satisfies the relation $eH>m$. We
learn that the relativistic vector theory is unstable at large
chemical potential whenever the non derivative vector self
interactions are tuned to be identical. This is precisely the
limit often used in literature when writing effective Lagrangians
that in QCD describe the $\rho$ vector field. In principle we can
still imagine to stabilize the potential in the gauge limit by
adding some higher order operators which seems unnatural. A more
natural solution to this instability is that the chemical
potential actually does not rise above the mass of the vectors
even if we increase the relative charge density. This phenomenon
is similar to what happens in the case of an ideal bose gas at
high chemical potential \cite{kapusta}.

 Interestingly by studying
the vector condensation phenomenon for strongly interacting
theories on the lattice at high isospin chemical potential we can
determine the best way of describing the ordinary vector
self-interactions at zero chemical potential.
\section{Dispersion Relations}
\label{dispersion} To determine the dispersion relations we
concentrate on the quadratic terms of the theory. {}For $\mu<m$
vectors do not condense and the only terms we need to consider are
the ones in eq.~(\ref{quadraticL}) which we report here for the
reader's convenience:
\begin{eqnarray}
{\cal{L}}_{\rm Quadratic} &=&\half \,\psi^\mu \, [ g_\mn
[\partial^2+m^2] -
\partial_\mu
\partial_\nu ] \, \psi^\nu + \left\{ \half \, {\phi^{\ast}}^\mu \, \left(
g_\mn [D^2 + m^2] - D_\mu D_\nu \right)\,  \phi^\nu + c.c.
\right\} \ .
\end{eqnarray}
These are the only terms we need for the $SU(2)$ theory with or
without the $\delta$ term. We note that in the covariant
derivative acting on $\phi$ it is hidden the negative $\mu^2$
square term appeared already in the potential term in the previous
section. The $\psi$ field is a standard massive vector field with
3 independent degrees of freedom since it satisfies the constraint
$m^2\partial_{\mu} \psi^{\mu}=0$. The associated dispersion
relations for the 3 physical components of the $\psi$ fields are:
\begin{eqnarray}
E_{\psi}= \sqrt{{\bf{p}}^2 + m^2} \ .
\end{eqnarray}
{}For the field $\phi$ we have the following equation of motion
\begin{eqnarray} \left( g_\mn [D^2 + m^2] - D_\mu D_\nu \right)\,
\phi^\nu = 0 \ ,\end{eqnarray} which by multiplying on the left by
$D_\mu$ leads to the constraint
$m^2 D_\nu\phi^\nu = 0$ reducing the number of physical degrees of
freedom for $\phi_{\mu}$ to three. Assuming the previous
constraint the equation of motion is clearly $[D^2 + m^2] \phi_\nu
= 0$, leading to the following dispersion relations
\cite{Sannino:2001fd}:
\begin{eqnarray}
E_{\phi^{\pm}} = \pm \mu +\sqrt{ {\bf{p}}^2 + m^2} \, .
\end{eqnarray}
Each of the ($\pm$) state corresponds to a positive and negative
charge under $U(1)$ and constitutes an $SO(3)$ vector with 3
independent degrees of freedom. Having reviewed the case $\mu<m$
we now analyze the dispersion relations of the system when the
vector condenses (i.e. $\mu>m$). The physical constraints are now
more involved.  Expanding our fields around the new vacuum of the
theory:
\begin{eqnarray}
\phi_{\mu} = \langle \phi_{\mu}\rangle+\widetilde{\phi}_{\mu} \ ,
\qquad \psi_{\mu}=\psi_{\mu} \ ,
\end{eqnarray}
new quadratic terms emerge depending on
$\langle\phi_{\mu}\rangle$. Since the specific properties of the
vacuum are very different for $\lambda^{\prime}\ge 0 $ or
$\lambda^{\prime}<0$ we consider these two cases separately.
\subsection{The polar phase dispersion relations}
\subsubsection{Dispersion relations for the $\psi$ field}
In the $\delta=0$ limit the quadratic terms of $\psi$ and $\phi$
decouple and we start analyzing the dispersion relations for the
field $\psi$. The most general Lagrangian term quadratic in the
fluctuation field $\psi$ is:
\begin{eqnarray}
\half \,\psi^\mu \, [ g_\mn [\partial^2+\bar{m}^2] -
\partial_\mu
\partial_\nu ] \, \psi^\nu + \frac{{\kappa}}{2} \psi^{\mu} n_{\mu}
n_{\nu} \psi^{\nu} \ ,
\end{eqnarray}
with $n_{\mu}= \langle \phi_{\mu} \rangle$ and
\begin{eqnarray}
\bar{m}^2=m^2 - 2\,{\lambda} n^{\mu}n_{\mu} \ , \qquad
{\kappa}=2\,\lambda^{\prime} \ .
\end{eqnarray}
Here we have already used the fact that $n_{\mu}$ is real.
Multiplying the associated equation of motions on the left by
$\partial^{\mu}$ the free field constraint is:
\begin{eqnarray}
\bar{m}^2 \partial_{\nu}\psi^{\nu}= -\kappa \,n\cdot \partial
\,n_{\nu}\psi^{\nu} \ . \label{c1}
\end{eqnarray}
A convenient way of dealing with this constraint is to split our
field in a component parallel and one orthogonal to the vacuum
expectation value via \footnote{I am indebted to W. Sch\"{a}fer
for suggesting this way of splitting the fields.}:
\begin{eqnarray}
\psi_{\mu}={\cal P}_{\mu\nu} \psi^{\nu}
+\frac{n_{\mu}n_{\nu}}{n^2} \psi^{\nu} \equiv {\psi_V}_{\mu} +
\frac{n_{\mu}}{\sqrt{-n^2}} {\psi_S} \ ,
\end{eqnarray}
with
\begin{eqnarray}
{\cal P}_{\mu\nu}=g^{\mu\nu} - \frac{n^{\mu}n^{\nu}}{n^2}\ ,
\qquad {\psi_V}_{\mu}&=&{\cal P}_{\mu\nu} \psi^{\nu} \ , \qquad
{\psi_S}=\frac{n_{\nu} \psi^{\nu}}{\sqrt{-n^2}} \ .
\end{eqnarray}
Clearly ${\psi_V}_{\mu}$ is transverse to the vev while ${\psi_S}$
is parallel. In terms of these new fields and using the constraint
in eq.~(\ref{c1}) the quadratic Lagrangian for $\psi$ as function
of a generic but real vev for $\phi$ is:
\begin{eqnarray}
\frac{1}{2}{\psi_V}^{\mu}\left( \partial^2 + \bar{m}^2\right)
{\psi_V}_{\mu} - \frac{1}{2}{\psi_S}\left[\left(\partial^2 +
\bar{m}^2\right) + n^2 \, \kappa - n^2\frac{\kappa^2}{\bar{m}^4}
(n\cdot \partial)^2\right]{\psi_S} \ .
\end{eqnarray}
{}For the 2-components transverse field ${\psi_V}$ we derive the
following dispersion relations:
\begin{eqnarray}
E_{\psi_V}^2={\bf p}^2 + \Delta^2_{\psi_V} \ , \qquad
\Delta^2_{\psi_V}=\bar{m}^2 \ ,
\end{eqnarray}
while for the (one component) longitudinal field ${\psi_S}$ we
have:
\begin{eqnarray}
E^2_{\psi_S}=p^2_{\perp}+v^2_{{\psi_S}\parallel}\,p^2_{\parallel}+
{\Delta}^2_{\psi_S}\ ,\quad
{\Delta}^2_{\psi_S}=\bar{m}^2+n^2\,\kappa \ ,\quad
v^2_{{\psi_S}\parallel} = 1 - \frac{\kappa}{\bar{m}^4} \langle
|\vec{n}| \rangle^4 \ .
\end{eqnarray}
Substituting the explicit expression for the vacuum we deduce:
\begin{eqnarray}
\Delta^2_{\psi_V}= \mu^2 + \frac{\lambda^{\prime}}{\lambda -
\lambda^{\prime}}\left(\mu^2 - m^2 \right) \ , \quad
\Delta^2_{\psi_S}= \mu^2 \ ,\quad
 v^2_{{\psi_S}\parallel} = 1 -
\frac{{\lambda^{\prime}}^2}{(\lambda -
\lambda^{\prime})^2}\frac{\mu^2 - m^2}{\bar{m}^2} \ ,
\end{eqnarray}
$p_{\parallel(\perp)}$ are the component of the momentum parallel
(perpendicular) to the vector condensate.

\subsubsection{Dispersion relations for the $\phi$ field} The
situation is more involved for the fluctuations of the complex
$\phi$ vector field. {}For these fields we write the general
quadratic Lagrangian in a 2 component formalism as follows:
\begin{eqnarray}
\frac{1}{2}{\chi^{\dagger}}^{\mu} \left[g_{\mu\nu}\left(\Delta^2 +
\overline{M} \right)- \Delta_{\mu}\Delta_{\nu} \right] \chi^{\nu}
+ \frac{1}{2}{\chi^{\dagger}}^{\mu} {\cal K}n_{\mu}
n_{\nu}\chi^{\nu} \ .
\end{eqnarray}
with
\begin{eqnarray} \chi = \left(%
\begin{array}{c}
  \widetilde{\phi} \\
  \widetilde{\phi}^{\ast} \\
\end{array}%
\right) \ , \quad \Delta_{\mu}=\left(%
\begin{array}{cc}
  D_{\mu} & 0 \\
  0 & {D_{\mu}}^{\ast}\\ \end{array} \right) \ , \quad \overline{M}=\left(%
\begin{array}{cc}
  \overline{m}^2 & r \\
  r^{\ast} & \overline{m}^2  \\ \end{array} \right)\ , \quad {\cal K}=\left(%
\begin{array}{cc}
 \beta & \alpha \\
  \alpha^{\ast} & \beta \\
\end{array}%
\right)
\end{eqnarray}
The equation of motion for $\chi$ is:
\begin{eqnarray}
 \left[g_{\mu\nu}\left(\Delta^2 + \overline{M} \right)-
\Delta_{\mu}\Delta_{\nu} \right] \chi^{\nu} + {\cal K}n_{\mu}
n_{\nu}\chi^{\nu} = 0 \ .
\end{eqnarray}
Multiplying on the left by $\Delta_{\mu}$ we obtain the following
physical condition:
\begin{eqnarray}
\overline{M} \Delta_{\nu} \chi^{\nu} = - \left( n\cdot \Delta
\right) {\cal K} \left( n\cdot \chi \right)\ . \label{chi}
\end{eqnarray}
We again split $\chi_{\mu}$ as follows:
\begin{eqnarray}
{{\Phi_V}}_{\mu} = {\cal  P}_{\mu \nu}\chi^{\nu}\ ,\qquad \
{\Phi_S}_{\mu} = \frac{n_{\nu}\chi^{\nu}}{\sqrt{-n^2}} \ ,
\end{eqnarray}
and using eq.~(\ref{chi}) the quadratic Lagrangian becomes:
\begin{eqnarray}
\frac{1}{2}{{\Phi}_{V}^{\mu}}^{\dagger}\left( \Delta^2 +
\overline{M}\right) {{\Phi}_{V}}_{\mu} -
\frac{1}{2}{{\Phi}_{S}}^{\dagger}\left[\left(\Delta^2 +
\overline{M}\right) + n^2 \, {\cal K} - n^2{{\cal K}^{\dagger}\,
{\overline{M}}^{-1} \, {\overline{M}}^{-1}\, \cal K } (n\cdot
\partial)^2\right]{\Phi}_{S} \ . \label{RealQuadratic}
\end{eqnarray}
In deriving the last equation we used the fact that the zeroth
component of $n_{\mu}$ vanishes. When the covariant derivative
acts on $\Phi$ is always $D$ while when acts on $\Phi^{\ast}$ is
always $ D^{\ast}$. We have an expression, formally, similar to
the one obtained for the neutral field $\psi$.

In terms of the coefficients of the effective Lagrangian we have
(after substituting the expression for the vev):
\begin{eqnarray}
\Delta^2 &=&\left(%
\begin{array}{cc}
  -E^2+{\bf p}^2+ 2\mu\, E - \mu^2 & 0  \\
   0&  -E^2+{\bf p}^2- 2\mu\, E - \mu^2 \\
\end{array}%
\right)\ , \quad
\overline{M} = \left(%
\begin{array}{cc}
  m^2 - \left(2\lambda - \lambda^{\prime} \right)\, n^2&  \lambda^{\prime} n^2 \\
  \lambda^{\prime} n^2 &  m^2 -  \left(2\lambda - \lambda^{\prime} \right)\, n^2 \\
\end{array}%
\right)\ ,  \nonumber \\ {\cal K} &=& -\left(%
\begin{array}{cc}
  2\lambda - 3\lambda^{\prime} & 2 \lambda - \lambda^{\prime} \\
  2 \lambda - \lambda^{\prime} &  2\lambda - 3\lambda^{\prime} \\
\end{array}%
\right) \ , \qquad n^2=-\frac{1}{2}\frac{\mu^2-m^2}{\lambda
-\lambda^{\prime}} \ .
\end{eqnarray}
Since in our case holds the relation:
\begin{eqnarray}
\overline{m}^2-\mu^2=|r| \ , \label{condition}
\end{eqnarray}
we have one zero gap vector field (with 2 independent physical
degrees of freedom) with respect to $SO(2)$ and one non zero gap
vector with
\begin{eqnarray}
\Delta^2_{\Phi_V^{+}}=2\left(\overline{m}^2 + \mu^2\right)=4\mu^2
+ \lambda^{\prime} \, \frac{\mu^2 - m^2}{\lambda -
\lambda^{\prime}} \ .
\end{eqnarray}
The full dispersion relations are (without enforcing yet the
condition (\ref{condition})):
\begin{eqnarray}
E^2_{\Phi_V^{\mp}}=\overline{m}^2+\mu^2 - {\bf p}^2 \mp
\sqrt{4\mu^2 {\bf p}^2 + r^2 + 4\overline{m}^2} \ ,
\end{eqnarray}
Expanding in momenta and enforcing eq.~(\ref{condition}):
\begin{eqnarray}
E^2_{\Phi_V^{-}} &=& \frac{\lambda^{\prime}}{\lambda
-\lambda^{\prime}} \frac{\mu^2 - m^2}{\Delta^2_{\Phi_V^+}} \, {\bf
p}^2 + 16\frac{\mu^4}{\Delta^6_{\Phi_V^+}}{\bf p}^4 + {\cal
O}(\mu^6\,\frac{{\bf p}^6}{\Delta^{10}_{\Phi_V^+}}) \ , \\
E^2_{\Phi_V^{+}} &=& \Delta^2_{\Phi_V^+} +\left(1 +
\frac{4\mu^2}{\Delta^2_{\Phi_V^+}}\right) {\bf p}^2 -
16\frac{\mu^4}{\Delta^6_{\Phi_V^+}}{\bf p}^4 + {\cal
O}(\mu^6\,\frac{{\bf p}^6}{\Delta^{10}_{\Phi_V^+}})  \ ,
\end{eqnarray}
Hence the propagation of the vector fields orthogonal to the vev
is isotropic.

\begin{figure}[hbtp]
\centering \leavevmode {\includegraphics[width=.48\textwidth,
height=.3\textwidth]{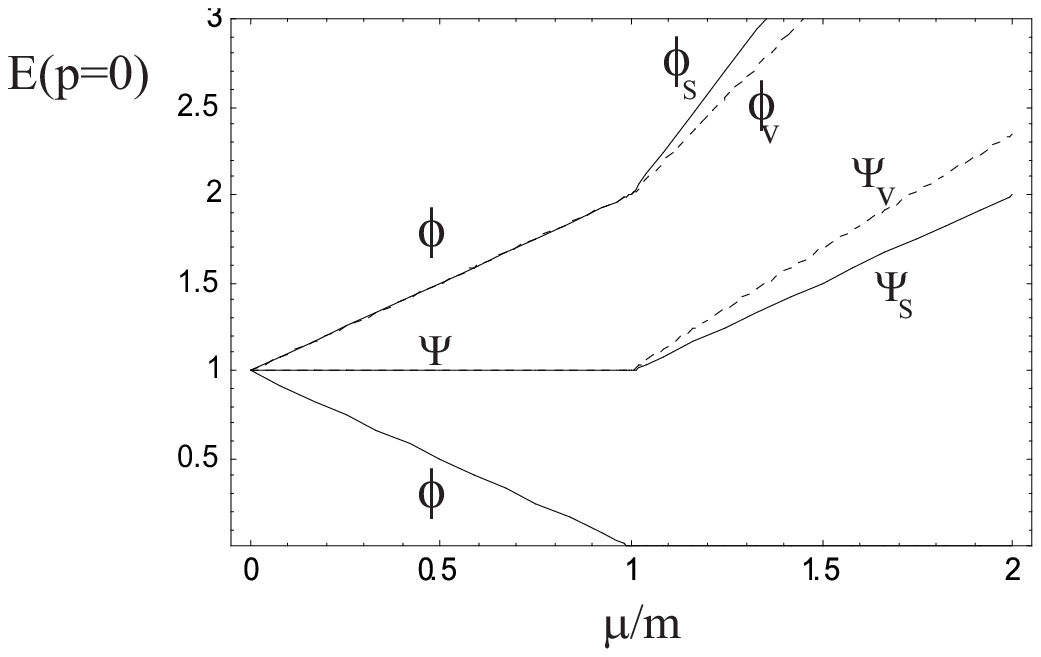}} \leavevmode
{\includegraphics[width=.48\textwidth,
height=.3\textwidth]{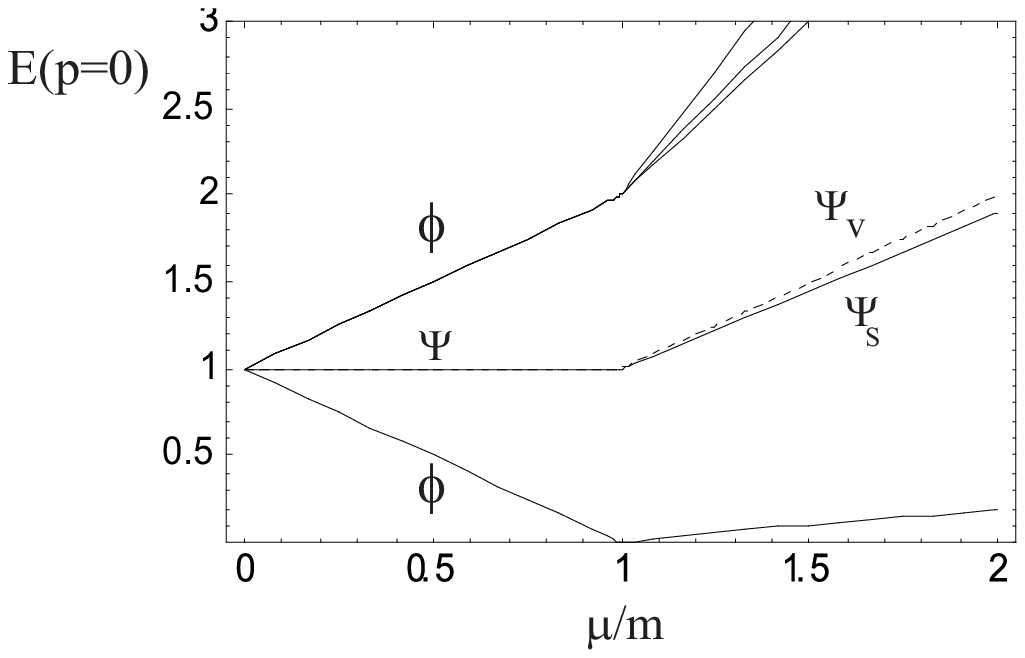}} \caption{We present the Gaps
($E(p=0)$) for the polar phase (left panel) and the apolar (right
panel) of the theory as function of the chemical potential in
units of $m$. Left Panel: Before condensation each line describes
three degenerate massive states. After condensation $\mu>m$ the
dashed lines represent two states while the solid lines represent
one state each except for the three gapless states. We used the
following values for the plot: $\lambda=1$,
$\lambda^{\prime}=0.33$. Right Panel: Before condensation each
line describes three degenerate massive states. After condensation
$\mu>m$ the dashed line represents two states ($\psi_V$) while the
solid lines one physical state each except for the two gapless
states. We used the following values for the plot: $\lambda=1$,
$\lambda^{\prime}=-0.3$.}\label{Figura2}
\end{figure}

We are now left to investigate ${\Phi_S}$. Using eq.
(\ref{RealQuadratic}) we get the following dispersion relations
for the gapless mode
\begin{eqnarray}
E_{{\Phi_S}^-}^2 &=& \frac{\mu^2-m^2}{3\mu^2 - m^2}
\left(p^2_{\perp} + {v}_{{\Phi_S}^-_{\parallel}}^2
p^2_{\parallel}\right) + \cdots \ ,
\nonumber \\
{v}_{{\Phi_S}^-_{\parallel}}^2&=& \mu^2\left(\lambda -
\lambda^{\prime} \right)\frac{\mu^2\lambda +
\lambda^{\prime}\left(\mu^2 - 2m^2\right)}{(\mu^2\lambda -m^2
\lambda^{\prime})^2} \ .
\end{eqnarray}
The massive mode dispersion relation is:
\begin{eqnarray}
E_{{\Phi_S}^+}^2 &=& 2\left(3\mu^2 - m^2 \right) + \frac{5\mu^2 -
m^2}{3\mu^2 - m^2} \left(p^2_{\perp} +
{v}_{{\Phi_S}^+_{\parallel}}^2 p^2_{\parallel}\right) + \cdots \ ,
\nonumber
\end{eqnarray}
with ${v}_{{\Phi_S}^+_{\parallel}}^2$ a lengthy but known
expression of the Lagrangian coefficients.
{}This completes the analytical study of the dispersion relations
for the case of the polar phase. We learn that the vector states
orthogonal to the condensate have isotropic dispersion relations
while the ones relative to the vector component in the direction
of the condensate are not isotropic. We also find that for
$\lambda^{\prime}=0$, as anticipated in the previous section, 2
gapless excitations have quadratic dispersion relations and hence
become type II goldstone bosons. This is related to the
enhancement of the global symmetry in the potential term. In
Fig.~\ref{Figura2} we plot the gaps as function of the chemical
potential for this phase in the left panel. Before condensation
each solid line corresponds to three physical states, after
condensation three gapless modes emerge and each dashed line
corresponds to two states while each solid one to a single state.

\subsection{The Apolar Phase Dispersion Relations and Gaps}
We also solved the physical constraint for the apolar case. In
this case the analytical analysis of the physical constraints and
of the dispersion relations is complicated by the fact that the
vacuum is complex. Since the computations are instructive but
technical we provided them in the last appendix and summarize here
the results. In this phase we have three broken generators but
only two gapless modes: a type I and a type II goldstone boson.
Since the unbroken generator is a linear combination of a rotation
and the generator for the internal $U(1)$ symmetry the gaps and
the dispersion relations lose the straightforward and nice
classification in doublets (i.e. vectors) and scalars with respect
to a standard rotation. The gaps in this phase, and for a specific
choice of the couplings, are displayed in the right panel of
Fig.~\ref{Figura2}. Before condensation each line describes three
degenerate massive states. After condensation $\mu>m$ the dashed
line ($\psi_V$) represents two states while the solid lines one
physical state each except for the two gapless states. We used the
following values for the plot: $\lambda=1$,
$\lambda^{\prime}=-0.3$. The dispersion relations too, in the
$\delta=0$ limit, display a more complex structure which does not
alter the gapless excitation structure and goldstone counting.

\subsection{Breaking the $Z_2$ symmetry.}

We now comment on what happens if we allow $Z_2$ symmetry breaking
terms such as the $\delta$ term. Before including the chemical
potential in the direction $T^3$ the absence of the $\delta$ term
prevents an odd number of vectors to be present in any vertex of
the theory. Since this term involves 3 fields it will not affect
the dispersion relations before condensation. After condensation
has taken place the vacuum structure (due to our ansatz) is also
unaffected by this term. Since the goldstone states are the
fluctuations around the vacuum in the direction of some of the
continuously broken symmetries their general properties are also
expected not to be disrupted. The $\delta$ term will, however,
change some of the details of the dispersion relations. The
possible quadratic terms in the fields emerging after condensation
will always mix the $\psi$ state with a $\phi$ one and their
effect will be investigated elsewhere.

\section{Physical Applications and Conclusions}
\label{conclusion}

We investigated the phase structure of the relativistic massive
vector condensation phenomenon due to a non zero chemical
potential associated to some of the global conserved charges of
the theory. The possible phase structure is very rich. Indeed
according to the value assumed by $\lambda^{\prime}$ we have three
independent phases. The polar phase with $\lambda^{\prime}$
positive is characterized by a real vacuum expectation value and 3
goldstone bosons of type I. The apolar phase for
$\lambda^{\prime}$ negative has a complex vector vacuum
expectation value spontaneously breaking CP. In this phase we have
one goldstone boson of type I and one of type II while still
breaking 3 continuous symmetries. The third phase has an enhanced
potential type symmetry and 3 goldstone bosons one of type I and
two of type II.

We also discovered that if we force the self interaction couplings
$\lambda$ and $\lambda^{\prime}$ to be identical, as predicted in
a Yang-Mills massive theory, our ansatz for the vacuum does not
lead to a stable minimum when increasing the chemical potential
above the mass of the vectors. A possible resolution of such an
instability is that the chemical potential can be at most as large
as the vector mass. This case it very similar to the Bose-Einstein
condensation phenomenon for an ideal bose gas at high chemical
potential. Interestingly the gauge coupling limit is often adopted
in literature to economically describe, for example, the QCD
composite vector field $\rho$. We hence suggest that lattice
studies at high isospin chemical potential in the vector channel
for QCD might be able to, indirectly, shed light on this sector of
the theory at zero chemical potential. More generally the hope is
that these studies might help understanding how to construct
consistent theories of interacting massive higher spin fields not
necessarily related to a gauge principle.

 We also developed a formalism which enabled us to
investigate the vacuum structure and dispersion relations in the
spontaneously broken phase of the theory. Our results are helpful
when trying to go beyond the classical and tree approximation. Our
present studies are readily applicable to a number of physical
phenomena of topical interest. {}For example in the framework of 2
color QCD at high baryon chemical potential vector condensation
has been predicted in \cite{Lenaghan:2001sd,{Sannino:2001fd}}.
Recent lattice studies \cite{Alles:2002st} seem to support it. The
present analysis while reinforcing the scenario of vector
condensation shows that we can have many different types of
condensations with very distinct signatures. {}Some details of
other possible physical applications are presented in
\cite{Sannino:2001fd}. The present analysis can be
straightforwardly extended to a general number of space dimensions
\cite{Sannino:2001fd} which may be useful for more exotic
scenarios related to the phenomenon of vector condensation
\cite{Li:2002iw}.

\acknowledgments I am very happy to thank P.H. Damgaard, H.B.
Nielsen and J. Schechter for valuable discussions, comments and
for critical reading of the manuscript. I wish to thank W.
Sh\"{a}fer for discussions, helpful comments and for collaborating
in the early stage of the project. K. Tuominen is thanked for
careful reading of the manuscript and for checking some of the
equations. I also acknowledge discussions with G.F.~Giudice, \'A.
M\'{o}csy, P. Olesen and K. Rummukainen.This work is supported by
the Marie--Curie fellowship under contract MCFI-2001-00181.

\appendix
\section{Lagrangian in cartesian components}
It is helpful to know also the different terms of the theory in
cartesian components. We start with the vector Lagrangian
presented and studied in the main test:
\begin{eqnarray}
{\cal
L}&=&-\frac{1}{4}F^a_{\mu\nu}F^{a{\mu\nu}}+\frac{m^2}{2}A_{\mu}^a
A^{a\mu} +\delta\, \epsilon^{abc}\partial{\mu} A_{a\nu} A^{\mu}_b
A^{\nu}_c  - \frac{\lambda}{4}\left(A^a_{\mu}A^{a{\mu}}\right)^2 +
 \frac{\lambda^{\prime}}{4} \left(A^a_{\mu}A^{a{\nu}}\right)^2 \ ,
 \end{eqnarray}
with $F_{\mu
\nu}^a=\partial_{\mu}A^a_{\nu}-\partial_{\nu}A^a_{\mu}$, $a=1,2,3$
and metric convention $\eta^{\mu \nu}={\rm diag}(+,-,-,-)$. Here,
$\delta$ is a real dimensionless coefficient, $m^2$ is the tree
level mass term and $\lambda$ and $\lambda^{\prime}$ are positive
dimensionless coefficients with $\lambda > \lambda^{\prime}$.

After including a nonzero chemical potential associated to a given
conserved charge - related to the generator (say $B$) - in the the
following way:
\begin{equation}
\partial_{\nu} A_{\rho} \rightarrow \partial_{\nu}A_{\rho} - i
\left[B_{\nu}\ ,A_{\rho}\right]\ ,
 \end{equation}
with $B_{\nu}=\mu \,\delta_{\nu 0} B\equiv V_{\nu} B$ where
$V=(\mu\ ,\vec{0})$. The vector kinetic term modifies according
to:
\begin{eqnarray}
{\rm Tr} \left[F_{\rho \nu} F^{\rho \nu}\right] &\rightarrow& {\rm
Tr} \left[F_{\rho \nu} F^{\rho \nu}\right] - 4i{\rm Tr}
\left[F_{\rho\nu}\left[B^{\rho},A^{\nu}\right]\right] - 2 {\rm Tr}
\left[\left[B_{\rho},A_{\nu}\right]\left[B^{\rho},A^{\nu}\right] -
\left[B_{\rho},A_{\nu}\right]\left[B^{\nu},A^{\rho}\right]\right]
\ .
\end{eqnarray}
The terms induced by $F_{\mu \nu}F^{\mu \nu}$, after integration
by parts, yields \cite{Lenaghan:2001sd}
\begin{eqnarray}
{\cal L}_{kinetic}&=&\frac{1}{2}A_{\rho}^a\left\{ \delta_{ab}
\left[g^{\rho\nu}\Box-\partial^{\rho}\partial^{\nu}\right]
-4i\gamma_{ab}\left[g^{\rho \nu}V\cdot \partial - \frac{V^{\rho}
\partial^{\nu} + V^{\nu}
\partial^{\rho} }{2} \right] +  2
\chi_{ab}\left[V\cdot V g^{\rho\nu}-V^{\rho}V^{\nu}\right]\right\}
A_{\nu}^b
\end{eqnarray}
with
\begin{equation}
\gamma_{ab}={\rm Tr}\left[T^a\left[B,T^b\right] \right]\ , \qquad
\chi_{ab}={\rm
Tr}\left[\left[B,T^a\right]\left[B,T^b\right]\right] \ .
\end{equation}
{}For $B=T^3$ we have \begin{equation}
\gamma_{ab}=-\frac{i}{2}\epsilon^{ab3} \ , \quad
\chi_{11}=\chi_{22}=-\frac{1}{2} \ , \quad \chi_{33}=0 \ .
\end{equation} The chemical potential induces a ``magnetic-type''
mass term for the vectors at tree-level.

The trilinear term with a single derivative is:
\begin{eqnarray}
{\cal L}_{\delta}= \delta \, \epsilon^{abc}\partial_{\mu} A_{a\nu}
A^{\mu}_b A^{\nu}_c + \delta \, \left(V\cdot A^b \right) A^{a\nu}
A_{\nu}^c
\left[\delta^{3b}\delta^{ac}-\delta^{3c}\delta^{ab}\right] \
.\end{eqnarray}.

\section{Cylindrical Coordinates}
Here we summarize all of the terms in the Lagrangian using the
cylindrical coordinates:
\begin{eqnarray}
\phi_\mu &=& {1 \over \sqrt{2}} ( A^1_\mu + i A^2_\mu )\ , \qquad
\phi^{\ast}_\mu = {1 \over \sqrt{2}} ( A^1_\mu - i A^2_\mu )\ ,
\qquad \psi_\mu = A^3_\mu \ ,
\end{eqnarray}
The quadratic, cubic and quartic terms - in the vector fields -
now reads:
\begin{eqnarray}
{\cal{L}}_{\rm quadratic} &=&\half \,\psi^\mu \, [ g_\mn
[\partial^2+m^2] -
\partial_\mu
\partial_\nu ] \, \psi^\nu + \left\{ \half \, {\phi^{\ast}}^\mu \, \left(
g_\mn [D^2 + m^2] - D_\mu D_\nu \right)\,  \phi^\nu + c.c.
\right\} \label{quadraticL}
\end{eqnarray}

\begin{eqnarray}
 {\cal L}_{cubic} &=& -i\, \delta \partial_{\mu} \psi_{\nu} \left({\phi^{\ast}}^{\mu}\cdot \phi^{\nu}\right) -i\,\delta\psi^{
 \nu} \left(\partial_{\mu}\phi^{\ast}_{\nu} \phi^{\mu}\right) -i\,\delta\psi^{
 \mu} \left(\partial_{\mu}\phi^{\ast}_{\nu}\phi^{\nu}\right) + {\rm c.c.}\nonumber \\ &&-2\delta\, V\cdot \psi \left(\phi^{\ast}\cdot
 \phi\right)- \delta\, \left[ V\cdot\phi^{\ast} \left(
 \phi\cdot \psi\right) + {\rm  c.c.} \right] \ .
 \end{eqnarray}

\begin{eqnarray}
{\cal L}_{4-vectors}=-
\frac{\lambda}{4}\left(A^a_{\mu}A^{a{\mu}}\right)^2 +
 \frac{\lambda^{\prime}}{4} \left(A^a_{\mu}A^{a{\nu}}\right)^2 &=&{\lambda^{\prime} - 2\lambda \over 2} (\phi^{\ast} \cdot \phi)^2 +
{\lambda^{\prime}\over 2} |(\phi \cdot \phi)|^2 +
{\lambda^{\prime} - \lambda \over 4} (\psi \cdot \psi)^2
\\ &&
-{\lambda} (\phi^{\ast} \cdot \phi)(\psi \cdot \psi) +
\lambda^{\prime} |(\psi \cdot \phi)|^2 \nonumber \\
\end{eqnarray}

\section{Dispersion relations for the apolar phase}
\label{Dispersion complex} In this case the vev is proportional to
the vector:
\begin{eqnarray}
n^{\mu}= \frac{1}{\sqrt{2}}\left(%
\begin{array}{c}
   0\\
   1 \\
   \imath \\
   0 \\
\end{array}%
\right) \ , \quad {\rm with} \quad n^2={n^{\ast}}^2=0\ , \quad
{\rm and}\quad  n\cdot n^{\ast} = -1 \ .
\end{eqnarray}
The quadratic Lagrangian term for the field $\psi$ takes the form:
\begin{eqnarray}
\half \,\psi^\mu \, [ g_\mn [\partial^2+\bar{m}^2] -
\partial_\mu
\partial_\nu ] \, \psi^\nu + \frac{{\kappa}}{2} \psi^{\mu} \left(n_{\mu}
n_{\nu}^{\ast} + n_{\mu}^{\ast} n_{\nu}\right)\psi^{\nu} \ ,
\end{eqnarray}
and the constraint equation reads:
\begin{eqnarray}
\bar{m}^2 \partial \cdot \psi = -{\kappa} \left[n\cdot
\partial \left(n^{\ast}\cdot\psi \right) + n^{\ast}\cdot
\partial \left( n\cdot\psi \right) \right]\ . \label{c2}
\end{eqnarray}
We define the following projectors
\begin{eqnarray}
{\cal P}_{\mu\nu}=g_{\mu\nu} - {\cal L}_{\mu\nu} \ , \qquad {\cal
L}_{\mu\nu}=\frac{n_{\mu}n^{\ast}_{\nu} +
n_{\nu}n^{\ast}_{\mu}}{n\cdot n^{\ast}} \ .
\end{eqnarray}
Now ${\cal L}_{\mu\nu}$ has two non zero components (i.e.
$\mu=\nu=x$ and $\mu=\nu=y$) and hence projects out two real
scalars while ${\cal P}_{\mu\nu}$ has non zero the temporal and
the zed component projecting out a vector which lives in $1+1$
dimensions (i.e. again a scalar field). It is convenient to split
$\psi$ as follows:
\begin{eqnarray}
\psi_{\mu}={\psi_{S}}_{\mu} + \frac{n_{\mu}}{n\cdot n^{\ast}}
{\psi_{D}} + \frac{n_{\mu}^{\ast}}{n\cdot n^{\ast}}
{\psi_D^{\ast}} \ , \qquad  {{\psi}_D}= n^{\ast} \cdot \psi \ ,
\end{eqnarray}
Using these fields plus the constraint the Lagrangian becomes:
\begin{eqnarray}
{\cal L}&=&\frac{1}{2}{\psi_S}_{\mu} \left(\partial^2 +
\bar{m}^2\right) {\psi_S}^{\mu} - \psi_D^{\ast} \left(\partial^2 +
\widetilde{m}^2 + \frac{\kappa^2}{ \bar{m}^4} (n\cdot
\partial)(n^{\ast} \cdot \partial) \right){\psi_D} \nonumber \\
&&-\frac{\kappa^2}{2\bar{m}^4} \left[{\psi_D}(n\cdot
\partial)^2 {\psi_D} + \psi_D^{\ast}(n^{\ast}\cdot
\partial)^2 \psi_D^{\ast}\right] \ ,
\end{eqnarray}
with
\begin{eqnarray}
\Delta^2_{\psi_D}\equiv\widetilde{m}^2=\bar{m}^2 - \kappa \ ,
\qquad \Delta^2_{\psi_S}\equiv \bar{m}^2 \ .
\end{eqnarray}
Diagonalizing the $D$ and $S$ sector independently we deduce the
following dispersion relations:
\begin{eqnarray}
E_{{\psi_D}^1}^2&=&\Delta^2_{\psi_D} + {\bf p}^2 \ , \qquad
E_{{\psi_D}^2}^2 = \Delta^2_{\psi_D} + v^2_{D^2}\left(p^2_x +
p^2_y\right) + p^2_z \ , \qquad  E_{\psi_S}^2 = \bar{m}^2 +{\bf
p}^2 \ ,
\end{eqnarray}
with \begin{eqnarray} v^2_{D^2} = 1 - \frac{\kappa^2}{\bar{m}^4} \
.\end{eqnarray} Substituting for the vev:
\begin{eqnarray}
\Delta^2_{\psi_S}=\bar{m}^2=\mu^2+\frac{\lambda^{\prime}}{2\lambda-\lambda^{\prime}}
\left(\mu^2 - m^2\right) \ , \quad
\kappa=\frac{\lambda^{\prime}}{2\lambda-\lambda^{\prime}}
\left(\mu^2 - m^2\right)\ ,\quad \Delta^2_{\psi_ D}=\mu^2 \ .
\end{eqnarray}
We see that in the $\psi$ sector of the theory when $\delta$ is
zero two states have isotropic dispersion relations but are not
degenerate. The two degenerate states have different momentum
dependence.

\subsection{The $\phi$ dispersion relations}
The general quadratic Lagrangian in a 2
component formalism and in the presence of the complex vev for
$\phi$ reads:
\begin{eqnarray}
\frac{1}{2}{\chi^{\dagger}}^{\mu} \left[g_{\mu\nu}\left(\Delta^2 +
\overline{M} \right)- \Delta_{\mu}\Delta_{\nu} \right] \chi^{\nu}
+ \frac{1}{2}{\chi^{\dagger}}^{\mu} {\cal K}_{\mu\nu}\chi^{\nu} \
.
\end{eqnarray}
with
\begin{eqnarray}
\overline{M}=\left(%
\begin{array}{cc}
  \mu^2 & 0 \\
  0 & \mu^2  \\ \end{array} \right)\ , \quad {\cal K}_{\mu\nu}=(m^2-\mu^2)\left(%
\begin{array}{cc}
  n_{\mu}n^{\ast}_{\nu}-\gamma\,n^{\ast}_{\mu}n_{\nu}& n_{\mu}n_{\nu} \\
 n^{\ast}_{\mu}n^{\ast}_{\nu}  &  n^{\ast}_{\mu}n_{\nu}-\gamma\,n_{\mu}n^{\ast}_{\nu} \\
\end{array}%
\right) \ , \quad {\gamma}=\frac{2\lambda^{\prime}}{2\lambda -
\lambda^{\prime}} \ .
\end{eqnarray}
Note that the matrix ${\cal K}_{\mu\nu}$ Lorentz structure is
solely determined by the vector vev. The quadratic equation of
motion leads to the following physical constraint:
\begin{eqnarray}
\overline{M} \Delta_{\nu} \chi^{\nu} = - \Delta^{\mu} {\cal
K}_{\mu\nu}  \chi^{\nu}  . \label{chi2}
\end{eqnarray}
We split $\chi$ as follows:
\begin{eqnarray}
\chi_{\mu}={\Phi_{S}}_{\mu} + \frac{n_{\mu}}{n\cdot n^{\ast}}
{\widetilde{\Phi}_{D}} + \frac{n_{\mu}^{\ast}}{n\cdot n^{\ast}}
{\Phi_D} \ , \qquad  {{\widetilde{\Phi}}_D}= n^{\ast} \cdot \chi \
,\quad {\Phi_D}= n \cdot \chi \ .
\end{eqnarray}
Now $\Phi_S$, $\Phi_D$ and $\widetilde{\Phi}_D$ represent three
independent (2-components) physical states. (i.e. using these
fields plus the constraint the quadratic term Lagrangian can be
compactly written as:
\begin{eqnarray}
\frac{1}{2}{{\Phi}_{S}^{\mu}}^{\dagger}\left( \Delta^2 +
\overline{M}\right) {{\Phi}_{S}}_{\mu} - \frac{1}{2}{{{\cal
T}}_{D}}^{\dagger}\left[\left(\Delta^2 + \overline{M}\right)\times
{\bf 1} + \frac{U}{\mu^4} - Q\right]{\cal T}_{D} \ ,
\label{ComplexQuadratic}
\end{eqnarray}
with
\begin{eqnarray}
{\cal T}_D=\left(%
\begin{array}{c}
  \widetilde{\Phi}_D \\
  \Phi_D \\
\end{array}%
\right)\ , \quad U=\left(%
\begin{array}{cc}
  {n^{\ast}}{\cal K}^{\dagger}\partial\,
  \partial{\cal K}{n} &
 { n^{\ast}}{\cal K}^{\dagger}\partial\,
  \partial{\cal K}{n^{\ast}} \\
  n{\cal K}^{\dagger}\partial\, \partial{\cal K}{n} &
  n{\cal K}^{\dagger}\partial\,
  \partial{\cal K}{n^{\ast}} \\
\end{array}%
\right) \ , \quad Q=\left(%
\begin{array}{cc}
  n^{\ast}{\cal K}n & n^{\ast}{\cal K}n^{\ast} \\
  n{\cal K} n  & n{\cal K}n^{\ast}  \\
\end{array}%
\right)
\end{eqnarray}
where ${\cal T}_D$ is a 4 column vector, $U$ and $Q$ are
four-dimensional matrices. $n{\cal K}^{\dagger}\partial\equiv
n^{\mu}{\cal K}^{\dagger}_{\mu\nu}\partial^{\nu}$ is a two
dimensional matrix.

The dispersion relations in the $S$ sector are now straightforward
and we deduce the following two states:
\begin{eqnarray} E_{GB-II}&=&\frac{p^2}{2\mu^2}+ {\cal O}(p^4) \ ,\\
E_{WBGB}&=& 2\mu^2 + \frac{p^2}{2\mu^2} + {\cal O}(p^4) \ .
\end{eqnarray}
The first state is a type II goldstone boson while the second
state is the would be goldstone boson e.g. the one which would
have been massless if we had no breaking of the Lorentz symmetry.
{}For the D sector the diagonalization can be performed
analytically when setting the momentum in the $x$ and $y$
direction to zero and we get:
\begin{eqnarray} E_{GB-I}^2&=& {v_z}^2\,p_z^2 + {\cal O}(p^4)\ ,\\
E_{\Phi_A}^2&=& 2(3\mu^2-m^2) + (2+v_z^2)\,p_z^2+{\cal O}(p^4) \ ,\\
E_{\Phi_B}^2&=& \Delta^2_{\Phi_B}+ v^2_{\Phi_{B}}\,p_z^2 + {\cal O}(p^4) \ ,\\
E_{\Phi_C}^2&=& \Delta^2_{\Phi_C} + v^2_{\Phi_{C}}\,{p_z^2} +
{\cal O}(p^4) \ .
\end{eqnarray}
This sector of the theory contains a goldstone boson of type I and
3 massive states.

In the following we summarize the nine physical gaps related to
this phase:
\begin{eqnarray}
\begin{array}{|cll||c|}
\hline &&&\\
\Delta^2_{GLess}&=&0\ , &{\rm 2~States} \\ &&& \\
\Delta^2_{\phi_{WBGB}}&=& 4\mu^2 & {\rm 1~State}
\\&&&\\ \hline &&&\\  \Delta^2_{\phi_A} &=& 2\,(3\mu^2 - m^2) &{\rm 1~State}
\\&&&\\
\Delta^2_{\phi_B} &=&
\frac{2}{(2\lambda-\lambda^{\prime})^2}(h1+\sqrt{h2}) &{\rm
1~State}
\\&&&\\
\Delta^2_{\phi_C}&=&\frac{2}{(2\lambda-\lambda^{\prime})^2}(h1-\sqrt{h2})
&{\rm 1~State} \label{imag-gaps}\\ \hline
\end{array}
\end{eqnarray}

\begin{eqnarray}
\begin{array}{|cll||c|}
\hline &&&\\
\Delta^2_{\psi_V} &=&\mu^2 & {\rm 2~States}\\&&&\\
\Delta^2_{\psi_S}&=&\mu^2+\frac{\lambda^{\prime}}{2\lambda-\lambda^{\prime}}(\mu^2-m^2)
&{\rm 1~State}\\\hline \end{array}\end{eqnarray}
 with:
\begin{eqnarray}
h1=m^2(2\lambda-\lambda^{\prime})\lambda^{\prime}+4\lambda^2\mu^2-
6\lambda\lambda^{\prime}\mu^2+2{\lambda^{\prime}}^2\mu^2, \qquad
h2=\mu^2\,(2\lambda-\lambda^{\prime})^3\,(2m^2\lambda^{\prime}+
(2\lambda-3\lambda^{\prime})\mu^2)\ .
\end{eqnarray}


\begin{references}




\bibitem{Linde:1979pr}
A.~D.~Linde,
Phys.\ Lett.\ B {\bf 86}, 39 (1979).



\bibitem{Ambjorn:1989gb}
J.~Ambjorn and P.~Olesen,
Phys.\ Lett.\ B {\bf 218}, 67 (1989) [Erratum-ibid.\ B {\bf 220},
659 (1989)], ibid. \ B {\bf 257}, 201 (1991), Nucl.\ Phys.\ B {\bf
330}, 193 (1990).


\bibitem{Manton:1979kb}
N.~S.~Manton,
Nucl.\ Phys.\ B {\bf 158}, 141 (1979).

\bibitem{Hosotani:1988bm}
Y.~Hosotani,
Annals Phys.\  {\bf 190}, 233 (1989).

\bibitem{Li:2002iw}
L.~F.~Li,
arXiv:hep-ph/0210063.


\bibitem{Brown:kk}
G.~E.~Brown and M.~Rho,
Phys.\ Rev.\ Lett.\  {\bf 66}, 2720 (1991).


\bibitem{Harada:2000kb}
M.~Harada and K.~Yamawaki,
Phys.\ Rev.\ Lett.\  {\bf 86}, 757 (2001) [arXiv:hep-ph/0010207].





\bibitem{BKY}  M.~Bando, T.~Kugo and K.~Yamawaki, Phys.~Rept. {\bf 164}, 217
(1988).



\bibitem{Pisarski:1999tv}
R.~D.~Pisarski and D.~H.~Rischke,
Phys.\ Rev.\ D {\bf 61}, 074017 (2000) [arXiv:nucl-th/9910056].



\bibitem{Buballa:2002wy}
M.~Buballa, J.~Hosek and M.~Oertel,
arXiv:hep-ph/0204275.



\bibitem{Lenaghan:2001sd}
J.~T.~Lenaghan, F.~Sannino and K.~Splittorff,
Phys.\ Rev.\ D {\bf 65}, 054002 (2002) [arXiv:hep-ph/0107099].



\bibitem{Sannino:2001fd}
F.~Sannino and W.~Schafer,
Phys.\ Lett.\ B {\bf 527}, 142 (2002) [arXiv:hep-ph/0111098].





\bibitem{Alles:2002st}
B.~Alles, M.~D'Elia, M.~P.~Lombardo and M.~Pepe,
arXiv:hep-lat/0210039. (See references therein for 2 color QCD at
high baryon chemical potential)


\bibitem{Muroya:2002ry}
S.~Muroya, A.~Nakamura and C.~Nonaka,
arXiv:hep-lat/0211010.

\bibitem{VBE}
D.M.~Stamper-Kurn, M.R.~Andrews, A.P.~Chikkatur, S.~Inouye,
H.-J.~Miesner, J.~Stenger, and W.~Ketterle, Phys.~Rev.~Lett.~{\bf
80}, 2027 (1998).


\bibitem{Volovik:2000mt}
G.~Volovik,
{\it J.\ Low.\ Temp.\ Phys.\ }  {\bf 121}, 357 (2000)
[arXiv:cond-mat/0005431] and references therein; U.~Leonhardt and
G.~E.~Volovik,
{\it JETP Lett. \ }  {\bf 72}, 46 (2000) [arXiv:cond-mat/0003428].





\bibitem{Nielsen}
H.B.~Nielsen and S.~Chadha,
Nucl.\ Phys.\ {\bf B105}, 445 (1976).



\bibitem{Schafer:2001bq}
T.~Schafer, D.~T.~Son, M.~A.~Stephanov, D.~Toublan and
J.~J.~Verbaarschot,
{\tt hep-ph/0108210}.
V.~A.~Miransky and I.~A.~Shovkovy,
{\tt hep-ph/0108178}.





\bibitem{KRS}  \"{O}.~Kaymakcalan, S.~Rajeev and J.~Schechter,
Phys.~Rev.~D{\bf 30}, 594 (1984); J.~Schechter, Phys.~Rev.~D{\bf
34}, 868 (1986); P.~Jain, R.~Johnson, Ulf-G.~Meissner, N.~W.~Park
and J.~Schechter, Phys.~Rev.~D{\bf 37}, 3252 (1988).

\bibitem{KS}  \"{O}.~Kaymakcalan and J.~Schechter, Phys.~Rev.~D{\bf 31},
1109 (1985).





\bibitem{ARS}  T.~Appelquist, P.S.~Rodrigues da Silva and F.~Sannino,
Phys.~Rev.~D{\bf 60}, 116007 (1999), {\tt hep-ph/9906555}.




\bibitem{DRS} Z.~Duan, P.S.~Rodrigues da Silva and F.~Sannino, Nucl.~Phys.~{\bf B} 592, 371
(2001), {\tt hep-ph/0001303}.


\bibitem{Ecker:1993de}
G.~Ecker, J.~Kambor and D.~Wyler,
Nucl.\ Phys.\ B {\bf 394}, 101 (1993).

\bibitem{kapusta}
J.~I.~Kapusta, {\rm Finite-temperature field theory}, Cambridge
Monographs on Mathematical Physics (1993).



\end{references}
\end{document}